\documentclass[preprint,12pt]{elsarticle}

\usepackage{amsmath,amssymb}
\usepackage{graphicx}
\graphicspath{{figures/}}
\usepackage{float}
\usepackage{placeins}
\usepackage{subfig}
\usepackage{booktabs}
\usepackage{array}
\usepackage{xcolor}
\usepackage[colorlinks=true,linkcolor=blue,citecolor=blue,urlcolor=blue]{hyperref}
\allowdisplaybreaks

\newcommand{\rev}[1]{#1}

\journal{Chinese Journal of Physics}

\begin{document}

\begin{frontmatter}

\title{Charged Penrose Extraction in RN--AdS Black Holes with Dark Energy: Ergosphere Geometry and Rigorous Efficiency Bounds}

\author[addr1]{Anirudh Pradhan}
\ead{pradhan.anirudh@gmail.com}
\author[addr2]{K. Ghaderi\corref{cor1}}
\ead{k.ghaderi@iau.ac.ir}
\author[addr3]{M. Zeyauddin}
\ead{uddin\_m@rcjy.edu.sa}
\cortext[cor1]{Corresponding author}

\address[addr1]{Centre for Cosmology, Astrophysics and Space Science (CCASS), GLA University, Mathura 281406, U.P., India}
\address[addr2]{Department of Physics, Mari.C., Islamic Azad University, Marivan, Iran}
\address[addr3]{Department of General Studies (Mathematics), Jubail Industrial College, Jubail 31961, Saudi Arabia}

\begin{abstract}
\rev{We develop a quantitative framework for charge assisted Penrose extraction in Reissner--Nordstr\"om Anti-de Sitter black holes surrounded by a Kiselev type anisotropic dark energy sector. The generalized electrostatic ergosphere is defined by the balance between redshift and electrostatic work. Under explicitly stated monotonicity assumptions on the lapse and electrostatic potential, we prove uniqueness of the exterior ergosphere boundary and of the turning point of negative energy trajectories. We derive a local upper bound on the escaping energy, formulate a finite radius reception criterion appropriate to AdS asymptotics, and map the allowed extraction domains in terms of the mass, charge, cosmological constant, Kiselev normalization, and state parameter.  We also introduce first order adiabatic discharge and accretion laws for slowly varying mass and charge, clarifying the regime in which the evolving horizon and ergosphere can be tracked consistently. The results provide controlled analytic and numerical diagnostics for charged Penrose extraction in RN--AdS backgrounds with a dark energy environment, while clearly separating local kinematic bounds from backreaction, radiative losses, and near extremal stability effects.}
\end{abstract}


\end{frontmatter}

\section{Introduction}

Direct imaging of black hole shadows in M87 and at the Galactic center has elevated black holes to firmly established astrophysical objects, while sharpening the need for precise theoretical frameworks in the strong field regime~[1-3]. Classical solutions of Einstein's equations typically contain spacetime singularities hidden behind horizons, with their genericity formalized by the singularity theorems~[4]. Although cosmic censorship posits that singularities remain cloaked, collapse studies have identified scenarios that end in naked singularities~[5-16], motivating effective descriptions that remain predictive at high curvature. One broad direction constructs regular spacetimes with finite curvature cores and complete geodesics~[17-23], including de Sitter core models in the spirit of Sakharov and Gliner, as realized by Bardeen and subsequent generalizations~[24-32], together with extensions in nonlinear electrodynamics and modified gravity, semiclassical settings, and associated phenomenology~[33-56]. At the same time, black holes serve as natural particle colliders~[57], and negative energy kinematics inside ergoregions enable energy extraction through the Penrose process~[58-62]. In charged geometries an electrostatic analogue operates: charged particles can occupy negative energy states within a generalized ergoregion set by the background potential, allowing charge assisted extraction in Reissner-Nordstr\"om  backgrounds~[63,64]. Hawking evaporation and gauge sector effects further suggest nontrivial charge dissipation channels and constraints on remnants~[65-69].

\rev{Embedding dark energy into black hole spacetimes offers an additional handle on long range structure without abandoning the successful tests of general relativity. A cosmological constant provides the simplest model of cosmic acceleration, while dynamical alternatives have been explored on theoretical and observational grounds~[70-75]. Scalar field quintessence is one such possibility~[76-78]. In the present work, however, the dark energy environment is not treated as a dynamical scalar field. We use the Kiselev construction as an effective anisotropic fluid model in a static spherical geometry~[79,80]. This distinction is important: the Kiselev source captures a controlled radial stress profile but should not be interpreted as a perfect fluid or as a complete cosmological quintessence sector. In parallel, a large literature has charted regular and hairy black hole families with improved near core behavior or additional fields, including noncommutative smearing of sources~[81-83], energy condition respecting hairy metrics~[84], eikonal shadow quasinormal mode correspondences~[85], conformal Killing constructions coupled to nonlinear electrodynamics and scalars with finite invariants~[86], Bardeen--Dirac stars in AdS with well defined light rings~[87], and signature change mechanisms~[88,89], with broader surveys in~[90-99]. These developments jointly motivate a systematic study of charged particle kinematics and extraction channels in Anti-de Sitter (AdS) backgrounds with a controllable effective dark energy sector.}

This work develops a coherent framework for charged Penrose extraction in a Reissner-Nordstr\"om Anti-de Sitter (RN-AdS) geometry surrounded by quintessence of Kiselev type. We characterize negative energy states for charged timelike particles and prove the uniqueness of their turning point outside the outer horizon, guaranteeing inevitable capture once a negative energy segment is entered and generalizing kinematic properties known from Kerr and RN backgrounds~[59-64] to the present setting. We then derive sharp local bounds for the charged Penrose process together with a practical reception criterion at a large but finite AdS radius, yielding clear parametric trends for the generalized electrostatic ergosphere and for extraction efficiency across $(M,Q,\Lambda,c,\omega_q)$. Finally, we formulate time dependent scenarios driven by monotone discharge, with or without concurrent accretion, in ingoing Eddington-Finkelstein gauge; from these we obtain evolution laws for the apparent horizon and for the generalized boundary delimiting negative energy motion, and construct feasibility maps that quantify how discharge and accretion reshape the window for extraction.

\rev{The analysis identifies parameter orderings that persist throughout the displayed domain: larger $M$ and $Q$, larger $c$, and the more negative value of $\omega_q$ widen the negative energy layer and raise the local efficiency bound, whereas the AdS barrier controls how far an escaping fragment can be received at finite radius. These conclusions are stated within the explicit monotonicity, test particle, and adiabatic assumptions given below. The framework complements prior studies of regular and hairy black holes~[17-37,44-56] and of the Penrose mechanism in rotating and charged geometries~[58-64], while situating charged extraction in an AdS setting with an effective anisotropic dark energy environment relevant to semiclassical and holographic model building~[100-106].}

Section~2 introduces the RN--AdS background with quintessence, formulates the charged motion setup, and delineates the generalized electrostatic ergoregion. Section~3 examines trajectories within the negative energy domain and establishes the uniqueness of the turning point together with inevitable capture. Section~4 develops the charged Penrose mechanism, deriving local efficiency bounds and a practical reception criterion. Section~5 presents the first order adiabatic evolution model, tracking the coupled motion of the horizon and the electrostatic boundary. Section~6 closes with conclusions and outlook.

\section{Charged motion and negative energy domain in RN-AdS spacetimes with quintessence}

\rev{A spherically symmetric charged black hole embedded in an Anti-de Sitter environment and surrounded by a Kiselev type anisotropic dark energy distribution is taken as the background geometry. The spacetime metric reads~[79,80]}
\begin{equation}
\mathrm{d}s^{2}=-f(r)\,\mathrm{d}t^{2}+f(r)^{-1}\mathrm{d}r^{2}+r^{2}\big(\mathrm{d}\theta^{2}+\sin^{2}\theta\,\mathrm{d}\phi^{2}\big),
\end{equation}
with metric function
\begin{equation}
f(r)=1-\frac{2M}{r}+\frac{Q^{2}}{r^{2}}-\frac{\Lambda r^{2}}{3}-\frac{c}{r^{3\omega_q+1}},
\qquad \Lambda<0,\ \ \omega_q\in(-1,-\tfrac13),\ \ c>0.
\end{equation}
The parameters $M$ and $Q$ represent the mass and electric charge of the central object, the constant $\Lambda$ fixes the AdS length scale $L_{\rm AdS}=\sqrt{-3/\Lambda}$, and the last contribution accounts for the Kiselev sector labeled by a state parameter $\omega_q$ and amplitude $c$. The associated effective matter content is taken in anisotropic form,
\begin{equation}
T^{\mu}{}_{\nu}{}^{(q)}=\mathrm{diag}(-\rho_q,\,-\rho_q,\,p_t,\,p_t), 
\qquad
p_t=\tfrac12(3\omega_q+1)\rho_q,
\end{equation}
with radius dependent density
\begin{equation}
\rho_q(r)=-\frac{c}{2}\,\frac{3\omega_q}{r^{3(1+\omega_q)}}>0,
\end{equation}
so that for the stated range of $\omega_q$ and $c>0$ one indeed obtains a positive energy density.

\rev{The source in Eqs.~(3)--(4) is an effective anisotropic stress tensor. It is used here as a phenomenological surrounding medium that modifies the radial lapse function. We do not assume that it is a perfect fluid, and we do not identify it with a full dynamical scalar field model of cosmological quintessence. The analysis therefore concerns charged test particle kinematics on this fixed effective geometry. Possible dynamical formation, perturbative stability of the anisotropic sector, and the response of the Kiselev medium to the fragments are outside the present approximation.}

The motion of a unit mass test particle with charge $q$ is restricted to the equatorial plane and evolves in a purely electrostatic background $A_\mu \mathrm{d}x^\mu=\Psi(r)\,\mathrm{d}t$. The potential is assumed to vanish at infinity and to decrease monotonically,
\begin{equation}
\Psi(\infty)=0, \qquad \Psi'(r)<0,
\end{equation}
which captures both the Maxwell choice $\Psi(r)=Q/r$ and a broad class of nonlinear electrodynamics models, while fixing a unique energy reference at spatial infinity. Minimal coupling together with the Killing fields $\partial_t$ and $\partial_\varphi$ gives the conserved specific energy and angular momentum~[107]
\begin{equation}
2\mathcal{L}=-f\,\dot t^2+f^{-1}\dot r^2+r^2\dot\varphi^2-2q\,\Psi(r)\,\dot t,
\qquad
E=f\,\dot t+q\,\Psi(r),\qquad L=r^2\dot\varphi,
\end{equation}
where overdots denote differentiation with respect to proper time. The timelike constraint $u^\mu u_\mu=-1$ produces the radial equation
\begin{equation}
\dot r^2=\Xi^2-f(r)\Big(1+\frac{L^2}{r^2}\Big),
\qquad
\Xi:=E-q\,\Psi(r),
\end{equation}
and forward evolution of coordinate time outside the outer event horizon $r_+$ requires
\begin{equation}
\dot t=\frac{\Xi}{f(r)}>0 \quad\Longrightarrow\quad \Xi>0\quad (r>r_+).
\end{equation}
Eliminating the velocities in favor of $E$ leads to
\begin{equation}
E=q\,\Psi(r)+\sqrt{\,f(r)\Big(1+\frac{L^2}{r^2}\Big)+\dot r^2\,},
\end{equation}
and the negative square root branch is discarded since it would violate the forward in time condition. At a fixed radius the smallest admissible energy occurs for $\dot r=0=L$ and equals
\begin{equation}
E_{\min}(r)=q\,\Psi(r)+\sqrt{f(r)},\qquad r>r_+.
\end{equation}

The existence of negative energy trajectories is tied to the sign of $q\,\Psi(r)$ relative to $\sqrt{f(r)}$. Adopting the convention $\Psi(r)>0$ and considering particles with charge opposite to the black hole potential, set $q=-|q|<0$. The boundary of the region where the conserved energy may become negative is then specified by [63]
\begin{equation}
|q|\,\Psi(r_{\mathrm E})=\sqrt{f(r_{\mathrm E})},
\end{equation}
which defines a generalized electrostatic ergosphere at $r=r_{\mathrm E}$. To analyze existence and uniqueness, introduce
\begin{equation}
G(r):=|q|\,\Psi(r)-\sqrt{f(r)},\qquad r>r_+.
\end{equation}
Monotonicity of the potential and positivity of $f'(r)$ outside the horizon imply
\begin{equation}
\begin{aligned}
G(r_+)&=|q|\,\Psi(r_+)>0, \\
\lim_{r\to\infty}G(r)&=-\infty, \\
G'(r)&=|q|\,\Psi'(r)-\frac{f'(r)}{2\sqrt{f(r)}}<0 .
\end{aligned}
\end{equation}
so $G$ decreases strictly from positive to negative values and crosses zero once and only once, which provides a unique $r_{\mathrm E}$ in the exterior domain. The location of this boundary varies smoothly with the magnitude of the particle charge. Implicit differentiation of the defining condition yields the sensitivity
\begin{equation}
\frac{\mathrm{d}r_{\mathrm E}}{\mathrm{d}|q|}=
\frac{\Psi(r_{\mathrm E})}{\displaystyle \frac{f'(r_{\mathrm E})}{2\sqrt{f(r_{\mathrm E})}}-|q|\,\Psi'(r_{\mathrm E})}>0,
\end{equation}
hence the negative energy domain expands outward as $|q|$ grows, while in the weak charge limit $r_{\mathrm E}\downarrow r_+$.

At the event horizon itself, the redshift factor vanishes and the conserved energy reduces to
\begin{equation}
E\big|_{r_+}=q\,\Psi(r_+),
\end{equation}
which shows that negative energy on the horizon is possible precisely when the particle charge is opposite in sign to the horizon potential. In the case of linear electrodynamics the potential equals $\Psi(r)=Q/r$ and the ergosphere condition becomes
\begin{equation}
\frac{|qQ|}{r_{\mathrm E}}=\sqrt{f(r_{\mathrm E})}.
\end{equation}
In the asymptotic AdS regime where $f(r)\sim r^{2}/L_{\rm AdS}^{2}$ one then finds $r_{\mathrm E}\sim (|qQ|L_{\rm AdS})^{1/2}$, which captures the outward drift of the negative energy boundary with increasing charge magnitude or with a longer AdS length scale. Dependence on the geometric and quintessence parameters follows from differentiating the defining relation with respect to any background variable $X\in\{M,Q,\Lambda,c,\omega_q\}$, which yields
\begin{equation}
\frac{\mathrm{d}r_{\mathrm E}}{\mathrm{d}X}
=
\frac{\displaystyle \frac{\partial_X f(r_{\mathrm E})}{2\sqrt{f(r_{\mathrm E})}} - |q|\ \partial_X\Psi(r_{\mathrm E})}
{\displaystyle \frac{f'(r_{\mathrm E})}{2\sqrt{f(r_{\mathrm E})}} - |q|\,\Psi'(r_{\mathrm E})},
\end{equation}
so that the shift of the ergosphere reflects both the variation of the metric function and any parametric dependence of the electrostatic potential. For instance, at fixed $\Psi$ one has $\partial_\Lambda f=-r^{2}/3$ and therefore $\partial_\Lambda \sqrt{f}=-r^{2}/(6\sqrt{f})<0$, indicating that a more negative cosmological constant pushes the ergosphere outward in the exterior region.

\rev{All numerical plots are produced in dimensionless geometric units, $G=c_{\rm light}=1$, with lengths scaled by the mass unit used in the metric. The parameters used in the figures are summarized in Table~\ref{tab:numerics}. Roots of $f(r)=0$, $G(r)=0$, $V_{\rm eff}(r)=0$, and the boundary curves in the feasibility maps were obtained by bracketed Brent root finding. The absolute and relative tolerances were set to $10^{-12}$; changing the tolerance to $10^{-10}$ does not change the plotted curves at visible resolution.}

\begin{table}[H]
\centering
\caption{\rev{Dimensionless numerical choices used in the plotted diagnostics.}}
\label{tab:numerics}
\scriptsize
\renewcommand{\arraystretch}{0.98}
\begin{tabular}{>{\raggedright\arraybackslash}p{0.20\textwidth}>{\raggedright\arraybackslash}p{0.25\textwidth}>{\raggedright\arraybackslash}p{0.43\textwidth}}
\toprule
\rev{Quantity} & \rev{Values used} & \rev{Role in figures} \\
\midrule
\rev{$M$} & \rev{$0.8,\ 1.2$} & \rev{Background mass in Figs.~1, 4--6} \\
\rev{$Q$} & \rev{$0.4,\ 0.6$} & \rev{Maxwell charge in Figs.~1, 4--6} \\
\rev{$\Lambda$} & \rev{$-0.05,\ -0.02$} & \rev{AdS curvature, $L_{\rm AdS}=\sqrt{-3/\Lambda}$} \\
\rev{$\omega_q$} & \rev{$-2/3,\ -1/2$} & \rev{Kiselev state parameter} \\
\rev{$c$} & \rev{$0.05,\ 0.10$} & \rev{Kiselev normalization} \\
\rev{$q$ in Fig.~1} & \rev{$0.01\le q\le0.70$} & \rev{Charge magnitude in the ergosphere map} \\
\rev{$|q|$ in Fig.~2} & \rev{$1\le |q|\le5$} & \rev{Continuous charge range for $V_{\rm eff}$} \\
\rev{$|q_1|$ in Figs.~3--4} & \rev{$1\le |q_1|\le5$} & \rev{Infalling fragment charge magnitude} \\
\rev{$E_0,q_2,L_2$} & \rev{$1,\ 0,\ 0$} & \rev{Reception and feasibility diagnostics} \\
\rev{$R_{\rm obs}$} & \rev{$10$} & \rev{Finite observer radius in Fig.~4} \\
\rev{Figs.~2--3 background} & \rev{$M=1.2,Q=0.6,\Lambda=-0.05$} & \rev{Uncrowded continuous charge maps} \\
\rev{$|q_1|_{\rm evol}$} & \rev{$1$} & \rev{Charge used to track $r_{\mathrm E}(v)$ in Figs.~5--6} \\
\rev{$\gamma$} & \rev{$5\times10^{-3}$} & \rev{Rate in $Q(v)=Q_{\rm ini}-\gamma v$} \\
\rev{$\mu$} & \rev{$5\times10^{-3}$} & \rev{Accretion rate in Scenario B} \\
\rev{Evolution interval} & \rev{$0\le v\le10$} & \rev{Figs.~5--6} \\
\bottomrule
\end{tabular}
\end{table}

\begin{figure}[t]
  \centering
  \includegraphics[width=\textwidth]{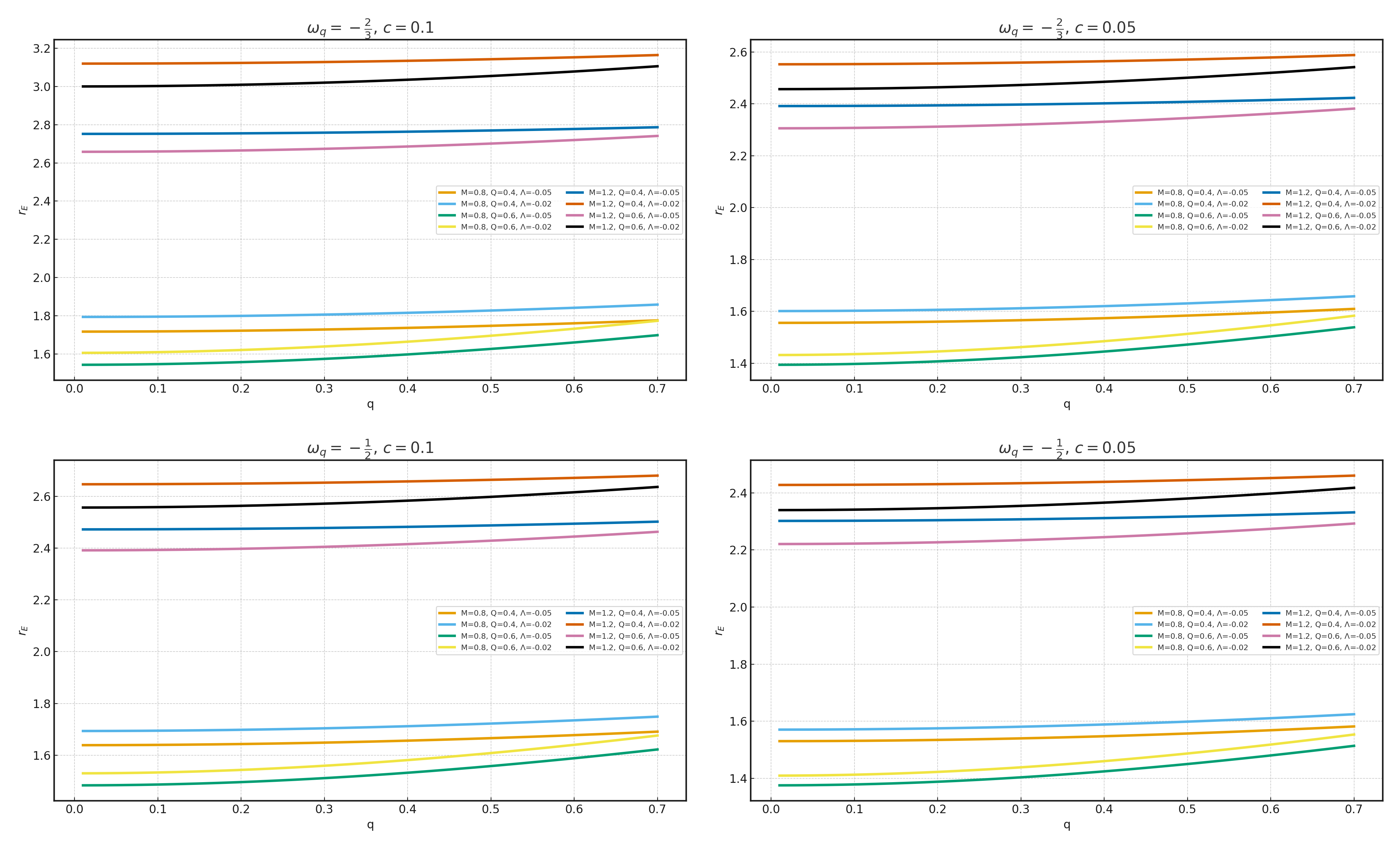}
  \caption{\rev{Generalized ergosphere radius $r_E$ versus charge magnitude $q$ in dimensionless units. Each panel fixes $(\omega_q,c)$, while curves correspond to $(M,Q,\Lambda)\in\{0.8,1.2\}\times\{0.4,0.6\}\times\{-0.05,-0.02\}$ as indicated by the legend. All roots are obtained by bracketed Brent solving of $|q|Q/r=\sqrt{f(r)}$.}}
  \label{fig:rEpanels}
\end{figure}

Figure~\ref{fig:rEpanels} displays the generalized ergosphere radius $r_E$ as a function of the positive charge parameter $q$ for four fixed background pairs $(\omega_q,c)$ and in each panel the curves correspond to the combinations $(M,Q,\Lambda)$.
The figure shows a clear monotonic growth of $r_E$ with $q$ in all panels. The slopes are small for very small $q$ and increase mildly as $q$ grows which is consistent with the fact that the electrostatic contribution that sustains negative energy trajectories strengthens with charge magnitude. For fixed $(\omega_q,c)$ the ordering across $(M,Q,\Lambda)$ is systematic. Larger $M$ produces larger $r_E$ at given $q$ which reflects the outward shift of the characteristic radii as the gravitational potential deepens. Making $\Lambda$ less negative shifts all curves downward while a more negative value of $\Lambda$ pushes them upward, hence the set with $\Lambda=-0.05$ sits above the set with $\Lambda=-0.02$ at fixed $(M,Q)$. The role of $Q$ is twofold. For smaller $q$ the difference between $Q=0.4$ and $Q=0.6$ is modest and $r_E$ is controlled mainly by the geometry. As $q$ grows the electrostatic term dominates the matching condition and the curves with larger $Q$ develop a slightly larger slope which is visible in the right half of each panel. There is no crossing between curves that share the same $(M,\Lambda)$ which indicates a robust ordering across $Q$ in the displayed range.

Comparing panels at fixed $\omega_q$ shows that increasing $c$ raises all profiles almost uniformly. This behavior is consistent with the extra background contribution from the quintessence sector which lowers the metric function in the exterior region and enlarges the domain in which the negative energy condition can be satisfied, thereby increasing $r_E$ for each $q$. The effect is more pronounced for the pairs with $M=1.2$ and with the more negative cosmological constant, leading to the highest curves in the figure. Changing $\omega_q$ from $-\tfrac{2}{3}$ to $-\tfrac{1}{2}$ reduces the overall level of the curves by a visible amount while preserving the same qualitative dependence on $(M,Q,\Lambda)$. This indicates that the state parameter controls a background scaling of the ergosphere radius without altering its monotonic response to the charge parameter.

The four panels together provide a compact parametric map of the generalized ergosphere radius. For all backgrounds with an outer horizon the function $r_E(q)$ is continuous, strictly increasing, and exhibits weak curvature over the interval $q\in[0.01,0.7]$. The families with larger $M$ and more negative $\Lambda$ occupy the upper envelope, while smaller $c$ and less negative $\omega_q$ shift the envelope downward. Since larger $r_E$ translates into a wider radial shell that supports negative energy kinematics, the figure implies that extraction scenarios are favored by larger mass, stronger AdS curvature, larger quintessence normalization, and by the more negative state parameter among the two values displayed.

\section{Trajectories of negative energy charged particles}
\label{sec:negE-trajectories}

We consider unit mass charged test particles moving on the equatorial plane in the electrostatic background $A_\mu \mathrm{d}x^\mu=\Psi(r)\,\mathrm{d}t$ with $\Psi(\infty)=0$ and $\Psi'(r)<0$. The conserved quantities are
\begin{equation}
E=f(r)\,\dot t+q\,\Psi(r),\qquad L=r^2\dot\varphi, 
\end{equation}
and forward evolution of coordinate time outside the outer horizon requires
\begin{equation}
\dot t=\frac{\Xi(r)}{f(r)}>0\quad\Longrightarrow\quad \Xi(r):=E-q\,\Psi(r)>0\quad (r>r_+),
\label{eq:forward_consistent}
\end{equation}
where $r_+$ denotes the largest positive root of $f(r)=0$. Timelike normalization yields the radial equation
\begin{equation}
\begin{aligned}
\dot r^2&=\Xi(r)^2-f(r)\Big(1+\frac{L^2}{r^2}\Big)
\equiv -\,V_{\mathrm{eff}}(r;E,L,q),\\
V_{\mathrm{eff}}(r)&:=f(r)\Big(1+\frac{L^2}{r^2}\Big)-\Xi(r)^2 .
\end{aligned}
\label{eq:Veff_consistent}
\end{equation}
In the AdS asymptotic domain one has $\Psi(r)\to 0$ and $\sqrt{f(r)}\sim r/L\to\infty$, hence $E_{\min}(r)=q\,\Psi(r)+\sqrt{f(r)}\to +\infty$ and a trajectory with $E<0$ cannot reach infinity. Such a worldline must exhibit at least one turning point $r_{\mathrm{tp}}>r_+$ with $V_{\mathrm{eff}}(r_{\mathrm{tp}})=0$ and $\dot r=0$. We show that this turning point is unique and that the motion then proceeds monotonically toward the horizon.

Differentiating \eqref{eq:Veff_consistent} gives
\begin{equation}
V_{\mathrm{eff}}'(r)=f'(r)\Big(1+\frac{L^2}{r^2}\Big)-2\,\frac{f(r)L^2}{r^3}-2\,\Xi(r)\,\Xi'(r),
\qquad
\Xi'(r)=-\,q\,\Psi'(r).
\label{eq:Veffprime_consistent}
\end{equation}
At a turning point one has $V_{\mathrm{eff}}(r_{\mathrm{tp}})=0$, which implies the bound
\begin{equation}
\frac{L^2}{r^2}\le \frac{\Xi^2-f}{f}\qquad (r\in[r_+,r_{\mathrm{tp}}]).
\label{eq:L_bound_consistent}
\end{equation}
Substituting \eqref{eq:L_bound_consistent} into \eqref{eq:Veffprime_consistent} yields
\begin{equation}
\begin{aligned}
V_{\mathrm{eff}}'(r)&\le f'(r)\Big(1+\frac{\Xi^2-f}{f}\Big)
-2\,\frac{f(r)}{r^3}\,r^2\,\frac{\Xi^2-f}{f}-2\,\Xi\,\Xi'\\
&= -\,\frac{f'(r)}{f(r)}\,\Xi^2+\frac{2}{r}\,(\Xi^2-f)-2\,\Xi\,\Xi' .
\end{aligned}
\label{eq:Veffprime_bound_consistent}
\end{equation}
Introduce the combinations
\begin{equation}
\Theta(r):=\Xi^2-f(r),\qquad \Upsilon(r):=\ln\!\Big(\frac{\Xi^2}{f(r)}\Big),
\label{eq:ThetaUpsilon_consistent}
\end{equation}
so that $\Upsilon'(r)=2\,\Xi'/\Xi - f'/f$ and \eqref{eq:Veffprime_bound_consistent} can be rewritten as
\begin{equation}
V_{\mathrm{eff}}'(r)\le \frac{\Xi^2}{2}\,\Upsilon'(r)+\frac{2}{r}\,\Theta(r).
\label{eq:Veffprime_key_consistent}
\end{equation}
For negative energy motion in the exterior domain, we adopt $\Psi(r)>0$ and $q=-|q|<0$, which together with $\Psi'(r)<0$ gives $\Xi'(r)=-q\,\Psi'(r)<0$. In RN-AdS with quintessence, within the parameter ranges and exterior intervals considered in this work, $f'(r)>0$ has been verified; therefore $\Upsilon'(r)=2\,\Xi'/\Xi-f'/f<0$ on $[r_+,r_{\mathrm{tp}}]$. Moreover, by construction of the generalized electrostatic ergosphere, $\Xi(r)<\sqrt{f(r)}$ along any negative energy trajectory outside $r_+$, so $\Theta(r)=\Xi^2-f<0$. Both terms on the right hand side of \eqref{eq:Veffprime_key_consistent} are then nonpositive and at least one is strictly negative on any interior subinterval, which implies
\begin{equation}
V_{\mathrm{eff}}'(r)<0\qquad \text{for all}\quad r\in[r_+,r_{\mathrm{tp}}].
\label{eq:Veffprime_negative_consistent}
\end{equation}
Since $V_{\mathrm{eff}}(r_+)=-\Xi(r_+)^2<0$ and $V_{\mathrm{eff}}(r_{\mathrm{tp}})=0$, the strict negativity \eqref{eq:Veffprime_negative_consistent} forbids any additional zero of $V_{\mathrm{eff}}$ in $[r_+,r_{\mathrm{tp}}]$. The turning point is therefore unique. With $\dot t>0$ from \eqref{eq:forward_consistent} and $\dot r^2=-V_{\mathrm{eff}}$, the motion for $r\le r_{\mathrm{tp}}$ is inward and the worldline inevitably crosses the event horizon in finite proper time.

\rev{The proof is conditional on three global properties in the exterior interval considered: (i) $f(r)>0$ for $r>r_+$, (ii) $f'(r)>0$ on the part of the exterior sampled by the negative energy trajectory, and (iii) a monotone electrostatic potential with $\Psi(\infty)=0$ and $\Psi'(r)<0$. These assumptions hold for the parameter windows used in Table~\ref{tab:numerics}; they were also checked numerically when the roots entering the figures were computed. If the conditions are relaxed, counterexamples can occur. For example, an exterior lapse with a local maximum or a nonmonotone nonlinear electrodynamics potential can make $G(r)$ or $V_{\rm eff}(r)$ develop more than one zero, so the uniqueness statement would no longer be global. For alternative matter models or nonlinear electrodynamics, the present argument carries over only after the corresponding $f'(r)$ and $\Psi'(r)$ signs have been verified in the exterior domain of interest.}

\begin{figure}[t]
  \centering
  \includegraphics[width=\textwidth]{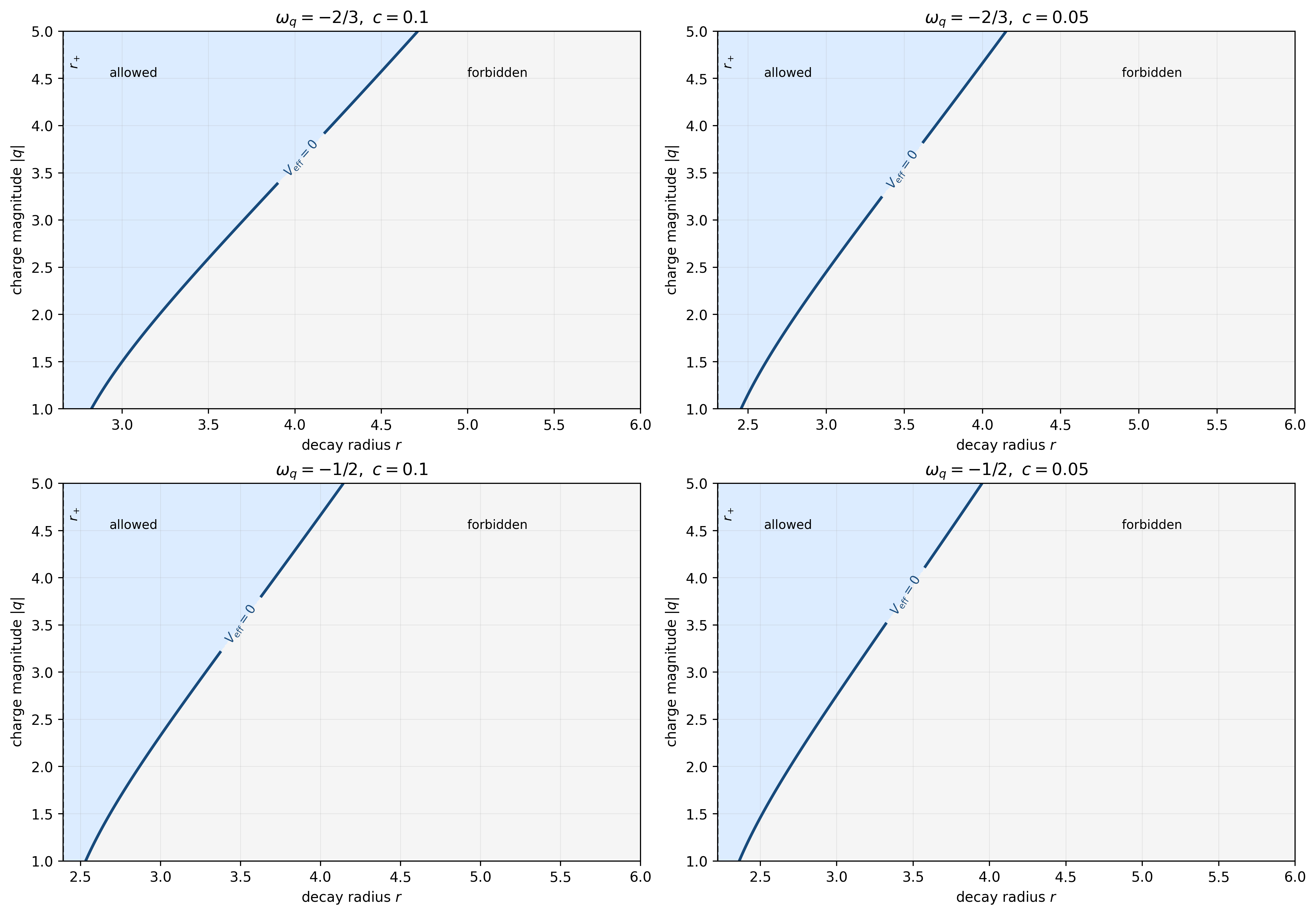}
  \caption{\rev{Effective potential domains for $E=0$ and $L=0$ in the representative background $M=1.2$, $Q=0.6$, and $\Lambda=-0.05$. Each panel fixes $(\omega_q,c)$. Blue shading marks the allowed region $V_{\rm eff}\le0$, grey shading marks the forbidden region $V_{\rm eff}>0$, the vertical dashed line is $r_+$, and the solid contour gives the marginal turning curve $V_{\rm eff}=0$.}}
  \label{fig:Veff2}
\end{figure}

\rev{Figure~\ref{fig:Veff2} has been drawn as a continuous charge domain. For the representative background $M=1.2$, $Q=0.6$, and $\Lambda=-0.05$, each panel fixes $(\omega_q,c)$ and displays the sign of $V_{\rm eff}(r;E=0,L=0,|q|)$ over $1\le |q|\le5$. The blue shaded region is the kinematically allowed set, $V_{\rm eff}\le0$, and the grey region is forbidden. The solid contour is the marginal turning curve $V_{\rm eff}=0$.}

\rev{The map makes the charge dependence transparent. Increasing $|q|$ shifts the marginal contour to larger radii because the electrostatic term $|q|Q/r$ more effectively compensates the redshift factor. Increasing $c$ or choosing the more negative state parameter $\omega_q=-2/3$ expands the blue allowed domain. These trends agree with the ergosphere radius in Fig.~\ref{fig:rEpanels}.}

\section{Penrose process in RN--AdS geometries with quintessence}
\label{sec:penrose}

A unit mass charged test particle labeled $0$ enters the generalized electrostatic ergoregion and undergoes a local decay
\[
0\to 1+2
\]
at a radius $r_\ast>r_+$ on the equatorial plane. Conservation of charge and four momentum at the decay point reads
\begin{equation}
q_0=q_1+q_2,\qquad p_0^{\mu}=p_1^{\mu}+p_2^{\mu},
\label{eq:penrose_cons}
\end{equation}
hence $E_0=E_1+E_2$, $p_0^r=p_1^r+p_2^r$, and $L_0=L_1+L_2$. Fragment $1$ is directed inward with negative conserved energy $E_1<0$, while fragment $2$ must propagate outward to a static detector at radius $R_{\mathrm{obs}}\ge r_\ast$.

To isolate the kinematic upper bound on the outgoing energy, evaluate the fragments \emph{at} $r_\ast$ with $\dot r_1=\dot r_2=0$ and $L_1=L_2=0$, and impose forward evolution of coordinate time, $\dot t=\Xi/f>0$ with
\[
\Xi:=E-q\,\Psi(r),\qquad \Psi(\infty)=0,\qquad \Psi'(r)<0,\qquad r>r_+.
\]
Under these choices the least energy compatible with forward time for fragment $1$ is
\begin{equation}
E_{1,\min}(r_\ast)=q_1\,\Psi(r_\ast)-\sqrt{f(r_\ast)}.
\label{eq:E1min_penrose}
\end{equation}
Negative values require $|q_1|\,\Psi(r_\ast)>\sqrt{f(r_\ast)}$ when $\Psi(r)>0$ and $q_1=-|q_1|<0$. Energy conservation then gives the largest energy that the escaping fragment can carry,
\begin{equation}
E_{2,\max}(r_\ast)=E_0-E_{1,\min}(r_\ast)=E_0+|q_1|\,\Psi(r_\ast)-\sqrt{f(r_\ast)},
\label{eq:E2max_penrose}
\end{equation}
and a local efficiency bound
\begin{equation}
\begin{aligned}
\eta(r_\ast)&:=\frac{E_2-E_0}{E_0}
\le \eta_{\max}(r_\ast)
:=\frac{|q_1|\,\Psi(r_\ast)-\sqrt{f(r_\ast)}}{E_0},\\
&\hspace{2.3cm}|q_1|\,\Psi(r_\ast)>\sqrt{f(r_\ast)} .
\end{aligned}
\label{eq:eta_penrose}
\end{equation}

The radial dependence of the bound follows from the same competition that defines the generalized ergoregion. Differentiating \eqref{eq:E2max_penrose} at fixed $q_1$ yields
\begin{equation}
\frac{\mathrm d E_{2,\max}}{\mathrm d r_\ast}=|q_1|\,\Psi'(r_\ast)-\frac{f'(r_\ast)}{2\sqrt{f(r_\ast)}}<0,
\label{eq:dE2dr}
\end{equation}
since $\Psi'(r_\ast)<0$ and $f'(r_\ast)>0$ for $r_\ast>r_+$. The gain is therefore higher for decays that occur deeper in the ergoregion, with the near horizon limit
\begin{equation}
E_{2,\max}\ \to\ E_0+|q_1|\,\Psi(r_+),\qquad
\eta_{\max}\ \to\ \frac{|q_1|\,\Psi(r_+)}{E_0}
\qquad (r_\ast\downarrow r_+),
\label{eq:nearH_penrose}
\end{equation}
which shows that the horizon value of the electrostatic potential fixes the ultimate local ceiling once a sufficiently negative fragment is available.

Sensitivity to background parameters at fixed $r_\ast$ follows directly from \eqref{eq:eta_penrose}. For any $X\in\{M,Q,\Lambda,c,\omega_q\}$ one has
\begin{equation}
\partial_X \eta_{\max}(r_\ast)=\frac{1}{E_0}\left(|q_1|\,\partial_X\Psi(r_\ast)-\frac{\partial_X f(r_\ast)}{2\sqrt{f(r_\ast)}}\right).
\label{eq:eta_param_sens}
\end{equation}
In the Maxwell case $\Psi(r)=Q/r$, $\partial_Q\Psi=1/r$ and $\partial_Q f=2Q/r^2$, whereas $\partial_\Lambda f=-r^2/3$ and $\partial_c f=-r^{-(3\omega_q+1)}$. Larger $M$ and larger $Q$ lower the lapse or raise the potential and thus increase $\eta_{\max}$, while making $\Lambda$ more negative raises the large radius barrier and suppresses the bound in the far region. Increasing $c$ or choosing the more negative state parameter moves the efficiency upward through the $f$ term, consistent with the expansion of the negative energy domain found earlier.

The charge split constrains how negative fragment $1$ can be. At fixed $q_0=q_1+q_2$ and given $r_\ast$, the escaping energy \eqref{eq:E2max_penrose} is maximized by taking $q_1$ as negative as allowed by the split and by $|q_1|\,\Psi(r_\ast)>\sqrt{f(r_\ast)}$. At the maximizing point one has $\Xi_1(r_\ast)=\sqrt{f(r_\ast)}+|q_1|\,\Psi(r_\ast)>0$, so forward time for fragment $1$ is automatically satisfied when $L_1=\dot r_1=0$ at $r_\ast$.

For the outward fragment the radial dynamics with the fixed notation is
\[
\Xi_2(r):=E_2-q_2\,\Psi(r),\qquad 
\dot r_2^{\,2}= \Xi_2(r)^2 - f(r)\Bigl(1+\frac{L_2^2}{r^2}\Bigr),
\]
and the forward in time condition requires $\dot t_2=\Xi_2/f>0$ outside $r_+$. An operational criterion for extraction is reception at a large but finite $R_{\rm obs}$ in the AdS exterior. Taking $L_2=0$ gives a conservative test. Under the same exterior monotonicity assumptions, $f'(r)>0$ and $\Psi'(r)<0$ for $r>r_+$, the function
\[
\mathcal{H}(r):=\Xi_2(r)^2-f(r)
\]
is minimized at $R_{\rm obs}$, hence the pointwise inequality
\begin{equation}
\bigl(E_2-q_2\,\Psi(R_{\rm obs})\bigr)^2 \ \ge\ f(R_{\rm obs})
\label{eq:escape_simple}
\end{equation}
is sufficient for escape, together with $\Xi_2(R_{\rm obs})>0$. Combining \eqref{eq:E2max_penrose} with \eqref{eq:escape_simple} yields the practical sufficient condition
\begin{equation}
E_0+|q_1|\,\Psi(r_\ast)-\sqrt{f(r_\ast)}\ \ge\ \sqrt{\,f(R_{\rm obs})\,}+q_2\,\Psi(R_{\rm obs}),
\label{eq:escape_combined}
\end{equation}
to be imposed alongside $\Xi_1(r_\ast)>0$ and $\Xi_2(R_{\rm obs})>0$. Whenever \eqref{eq:escape_combined} holds, fragment $2$ extracts net energy ($E_2>E_0$ because $E_1<0$) and reaches the detector, while fragment $1$ inevitably crosses the horizon.

\rev{The finite radius prescription is essential in AdS because a massive particle does not generically reach the conformal boundary with finite conserved energy. The value $R_{\rm obs}=10$ used in the feasibility plots is therefore not intended as an asymptotic infinity. It is a fiducial detector radius in the exterior region, large compared with the horizons in Table~\ref{tab:numerics} but finite compared with the AdS scale: $R_{\rm obs}/L_{\rm AdS}\simeq1.29$ for $\Lambda=-0.05$ and $R_{\rm obs}/L_{\rm AdS}\simeq0.82$ for $\Lambda=-0.02$. Restoring dimensions, one length unit corresponds to $GM_{\rm phys}/c_{\rm light}^{2}$; for a $10M_\odot$ black hole, $R_{\rm obs}=10$ would correspond to about $148$ km. The AdS and Kiselev parameters used here should therefore be read as controlled theoretical scales rather than as a direct fit to the observed cosmological constant. The observational utility of the calculation is to provide finite radius diagnostics and parameter orderings that can be compared with more realistic non AdS or rotating extensions.}

\rev{Figure~3 has also been drawn as a continuous charge map and shows $\eta_{\max}(r_\ast,|q_1|)$ over the full interval $1\le |q_1|\le5$ for the same representative background used in Fig.~\ref{fig:Veff2}. The colored domain is the region where the local gain is positive. The white contour is $\eta_{\max}=0$, and the blank region is locally inadmissible. This directly displays how the positive efficiency window grows with increasing negative fragment charge and contracts with increasing decay radius.}

\rev{Figure~4 retains the full multi background feasibility survey for the combined extraction plus reception requirement. Thresholds remain smallest close to the horizon and grow with radius; larger $M$ and $Q$ enlarge the domains, a more negative cosmological constant modifies the far region reach through the AdS barrier, and larger $c$ or the more negative $\omega_q$ expand the admissible domains. Together, the continuous maps and the feasibility plot provide a clearer parametric picture of the charged Penrose mechanism.}

\begin{figure}[t]
  \centering
   \includegraphics[width=\textwidth]{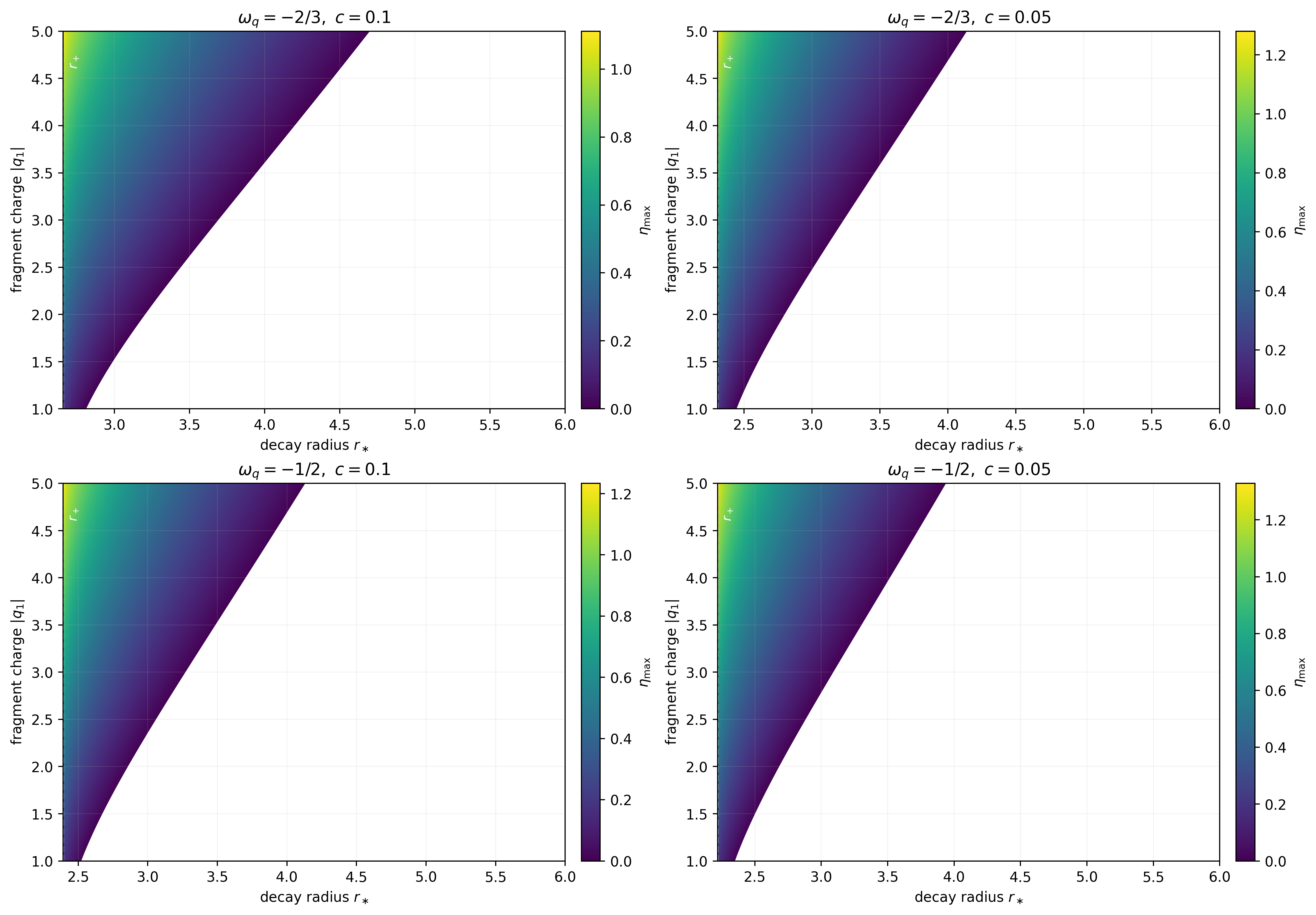}
  \caption{\rev{Continuous map of the maximum local efficiency $\eta_{\max}(r_\ast,|q_1|)$ for $E_0=1$ in the representative background $M=1.2$, $Q=0.6$, and $\Lambda=-0.05$. Each panel fixes $(\omega_q,c)$ and covers $1\le |q_1|\le5$. Colored regions have $\eta_{\max}>0$, the white contour marks $\eta_{\max}=0$, and the vertical dashed line is the outer horizon $r_+$.}}
  \label{fig:eta_max_2x2}
\end{figure}

\begin{figure}[t]
  \centering
  \includegraphics[width=\textwidth]{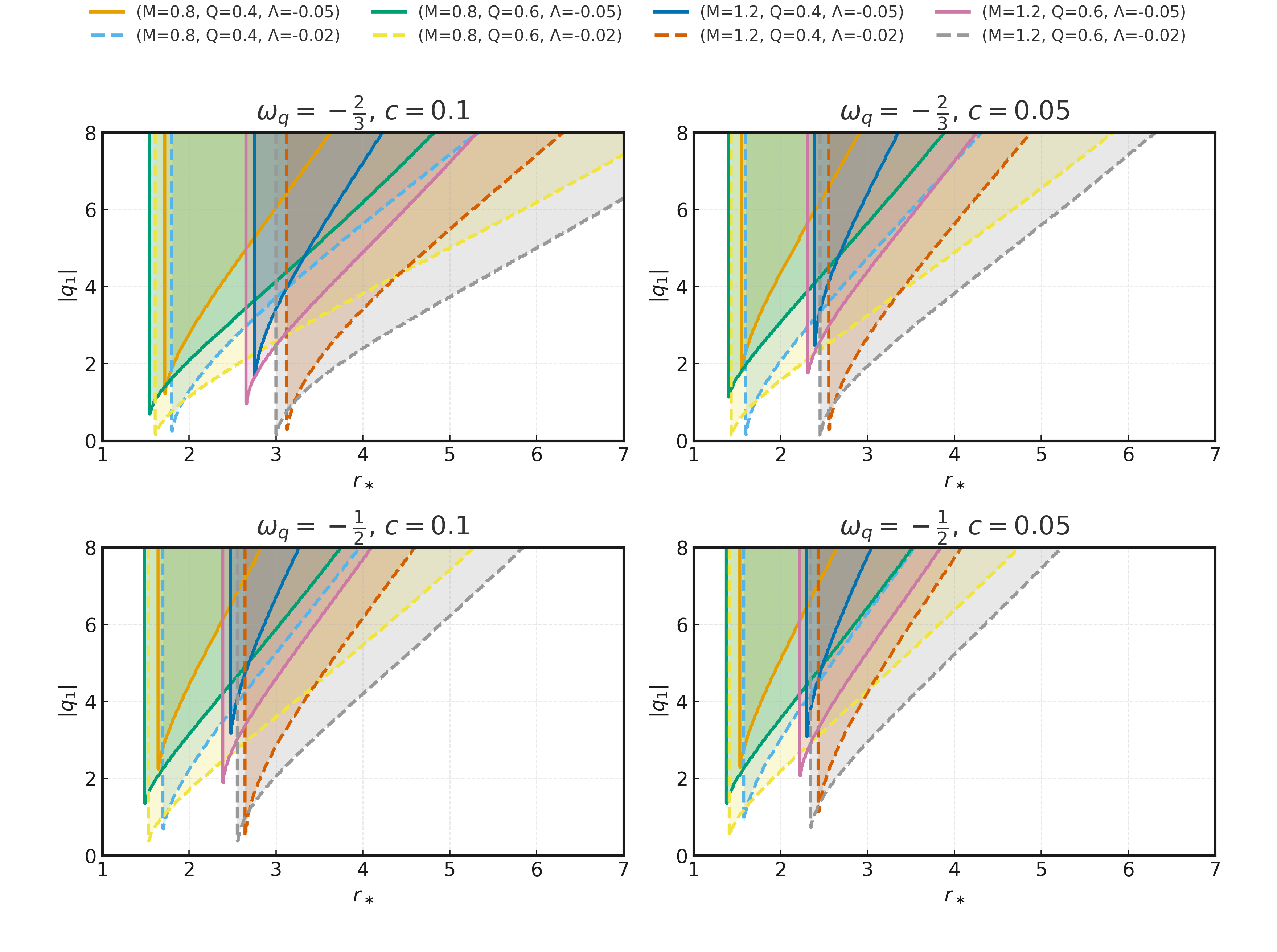}
  \caption{\rev{Feasibility domains in the $(r_\ast,|q_{1}|)$ plane for the combined extraction plus finite radius reception requirement with $E_{0}=1$, $q_{2}=0$, $L_{2}=0$, and $R_{\rm obs}=10$. Panel titles indicate $(\omega_q,c)$; the legend maps color and line style to $(M,Q,\Lambda)$. Shaded regions mark $|q_{1}|\ge |q_{1}|_{\min}(r_\ast)$, bounded by bold contours obtained by numeric root finding.}}
  \label{fig:feasibility_2x2}
\end{figure}

Event horizons are present whenever the lapse function admits at least one positive root. The boundary between black hole and horizonless geometries is the extremal locus where a double root develops at some $r=r_{\rm ext}>0$:
\begin{equation}
f(r_{\rm ext})=0,\qquad f'(r_{\rm ext})=0.
\label{eq:ext_system}
\end{equation}

Solving $f=0$ for the mass parameter gives a convenient parametric form
\begin{equation}
M(r;\,Q,\Lambda,c,\omega_q)
=\frac{r}{2}\left[\,1+\frac{Q^{2}}{r^{2}}-\frac{\Lambda r^{2}}{3}-\frac{c}{r^{3\omega_q+1}}\,\right].
\label{eq:M_of_r}
\end{equation}
Imposing $f'(r)=0$ at the same $r$ determines the extremal charge for that radius,
\begin{equation}
Q_{\rm ext}^{2}(r;\Lambda,c,\omega_q)
= r^{2}-\Lambda r^{4}+3c\,\omega_q\,r^{\,1-3\omega_q},
\label{eq:Qext_of_r}
\end{equation}
and substituting \eqref{eq:Qext_of_r} into \eqref{eq:M_of_r} yields the extremal mass in closed form,
\begin{equation}
M_{\rm ext}(r;\Lambda,c,\omega_q)
= r-\frac{2}{3}\,\Lambda r^{3}+\frac{(3\omega_q-1)\,c}{2\,r^{\,3\omega_q}}.
\label{eq:Mext_of_r}
\end{equation}
Equations \eqref{eq:Qext_of_r}-\eqref{eq:Mext_of_r} provide a single parameter representation of the threshold surface in the $(M,Q,\Lambda,c,\omega_q)$ space: for a given choice of $(\Lambda,c,\omega_q)$, every $r>0$ that satisfies $Q_{\rm ext}^{2}(r)\ge 0$ defines an admissible extremal point $(M_{\rm ext}(r),\,Q_{\rm ext}(r))$. Black holes exist precisely when there is a root of $f$ in the exterior, equivalently when the parameters lie above the extremal curve,
\begin{equation}
\begin{aligned}
M&\ge M_{\rm ext}(r_{\rm ext};\Lambda,c,\omega_q),\\
Q^{2}&\le Q_{\rm ext}^{2}(r_{\rm ext};\Lambda,c,\omega_q),
\qquad r_{\rm ext}>0 .
\end{aligned}
\label{eq:BH_region_condition}
\end{equation}
It is often convenient to eliminate $M$ and parametrize the boundary in the $(Q,\Lambda)$–plane at fixed $(c,\omega_q,r)$:
\begin{equation}
\begin{aligned}
\Lambda_{\rm ext}(r;\,Q,c,\omega_q)
&=\frac{r^{2}-Q^{2}+3c\,\omega_q\,r^{\,1-3\omega_q}}{r^{4}},\\
M_{\rm ext}(r;\,Q,c,\omega_q)
&=\frac{r}{2}\left[1+\frac{Q^{2}}{r^{2}}
-\frac{\Lambda_{\rm ext} r^{2}}{3}-\frac{c}{r^{3\omega_q+1}}\right].
\end{aligned}
\label{eq:Lamext_param}
\end{equation}

The condition $Q_{\rm ext}^{2}(r)\ge 0$ imposes a nontrivial constraint on the admissible extremal radii. Since $\omega_q\in(-1,-\tfrac{1}{3})$, the term $3c\,\omega_q\,r^{\,1-3\omega_q}$ in \eqref{eq:Qext_of_r} is negative and tightens the bound relative to the RN-AdS case; physically, the quintessence sector tends to shrink the region where a charged horizon can exist at a given $(Q,\Lambda)$. The mass profile \eqref{eq:Mext_of_r} shows three additive contributions: the linear term $r$ and the AdS term $-\tfrac{2}{3}\Lambda r^{3}$ mimic RN-AdS, while the quintessence piece $\tfrac{(3\omega_q-1)c}{2}\,r^{-3\omega_q}$ lowers $M_{\rm ext}$ because $(3\omega_q-1)<0$ in the range of interest. Consequently, for fixed $(\Lambda,c,\omega_q)$ the inclusion of quintessence generally enlarges the black hole region in the $(M,Q)$ plane when $r_{\rm ext}$ is held fixed, even though it reduces the allowed extremal charge through \eqref{eq:Qext_of_r}. The second derivative $f''(r)$ evaluated at \eqref{eq:ext_system} is positive on the admissible branch that yields the outer horizon, so the double root corresponds to a local minimum of $f$, as expected for extremal black holes. These features have been checked analytically from \eqref{eq:Qext_of_r}-\eqref{eq:Mext_of_r} and numerically across the parameter windows used in our figures.

\rev{The kinematic ceiling \eqref{eq:E2max_penrose} and the reception test \eqref{eq:escape_simple} are insensitive to the details of the extremality construction beyond the existence of an outer horizon. Approaching extremality ($r_+\to r_{\rm ext}$) drives $\sqrt{f(r_\ast)}\downarrow0$ for decay radii $r_\ast$ near $r_+$ and enhances the local factor $|q_1|\Psi(r_\ast)-\sqrt{f(r_\ast)}$. This is a local test particle statement, not a claim that an exactly extremal or fully backreacting system realizes arbitrarily large extraction. Near extremal geometries are known to be sensitive to horizon perturbations, including Aretakis type instabilities and related electromagnetic gravitational perturbations~[108,109]. Quantum discharge, pair creation, radiative losses, and the stress energy of the fragments can also regulate the idealized bound. Thus the enhancement reported here should be understood as the behavior of the first order kinematic ceiling within the nonextremal parameter region where the test particle and adiabatic approximations remain valid.}

\section{Illustrative charge evaporation dynamics in RN--AdS with quintessence}
\label{sec:evaporation-RNAdS-q}

Ingoing Eddington-Finkelstein coordinates $(v,r,\vartheta,\varphi)$ provide a convenient framework to describe slow evolution driven by external fluxes. We take
\begin{equation}
\mathrm{d}s^{2}=-f(r;M,Q,\Lambda,c,\omega_q)\,\mathrm{d}v^{2}+2\,\mathrm{d}v\,\mathrm{d}r
+r^{2}\big(\mathrm{d}\vartheta^{2}+\sin^{2}\!\vartheta\,\mathrm{d}\varphi^{2}\big),
\label{eq:EF-RNAdS-q}
\end{equation}
and allow a controlled $v$ dependence in the parameters that are directly tied to charged extraction: a slowly varying mass $M=M(v)$ and a slowly varying Maxwell charge $Q=Q(v)$ that sources the electrostatic potential seen by the probe. In keeping with the notation of the previous sections, we specialize the background potential to the Maxwell profile
\begin{equation}
A_\mu\,\mathrm{d}x^\mu=\Psi(v,r)\,\mathrm{d}v,\qquad
\Psi(v,r)=\frac{Q(v)}{r},
\label{eq:psi-Maxwell-time}
\end{equation}
so that charge evaporation corresponds to $\dot Q(v)\le 0$. Quintessence parameters $(c,\omega_q)$ and the cosmological constant $\Lambda<0$ are regarded as fixed external data.

\rev{The following evolution equations are not a full solution of the coupled Einstein--Maxwell--Kiselev system with arbitrary time dependence. They are first order adiabatic tracking relations for a nonextremal geometry whose parameters vary slowly compared with the instantaneous near horizon scale. For every Penrose fragment $i$ and for the slowly evolving background we require}
\begin{equation}
\begin{aligned}
\rev{\frac{E_i}{M}}&\rev{\ll1,}
\rev{\frac{|q_i|}{|Q|}}&\rev{\ll1,}
\rev{\epsilon_{\rm ad}}\rev{\equiv
\max\left\{
\kappa_+^{-1}\left|\frac{\dot M}{M}\right|,
\kappa_+^{-1}\left|\frac{\dot Q}{Q}\right|,
\kappa_+^{-2}\left|\frac{\ddot M}{M}\right|,
\kappa_+^{-2}\left|\frac{\ddot Q}{Q}\right|
\right\}\ll1 .}
\end{aligned}
\label{eq:adiabatic-validity}
\end{equation}
\rev{where $\kappa_+=\tfrac12\partial_r f|_{r_+}$ is the instantaneous nonextremal surface gravity. Accordingly, Eqs.~\eqref{eq:rplus-evolution}--\eqref{eq:deta-dv} retain only the terms linear in $\dot M$ and $\dot Q$; corrections of order $\epsilon_{\rm ad}^2$, $(\dot M)^2$, $(\dot Q)^2$, $\dot M\dot Q$, and higher derivative terms are outside the controlled approximation. The fragment stress energy, radiation reaction, and the dynamical response of the Kiselev sector are not included. For the linear laws used below, $\ddot M=\ddot Q=0$ and the constants $\gamma$ and $\mu$ listed in Table~\ref{tab:numerics} are chosen as small dimensionless rates.}

\rev{For the metric \eqref{eq:EF-RNAdS-q} with $M(v)$ and $Q(v)$, the mixed flux component along the ingoing null direction is fixed by the $v$ derivative of the lapse. With $\partial_v f=-2\dot M/r+2Q\dot Q/r^2$, the first order effective influx can be written as}
\begin{equation}
\rev{T^{\;r}{}_{v}=\frac{1}{4\pi r^2}\left(\dot M(v)-\frac{Q(v)\dot Q(v)}{r}\right)
=T^{\;r}{}_{v}\Big|_{\rm geom}+T^{\;r}{}_{v}\Big|_{\rm em},}
\label{eq:Tvr-split}
\end{equation}
\begin{equation}
\rev{T^{\;r}{}_{v}\Big|_{\rm geom}=\frac{\dot M(v)}{4\pi r^{2}},
\qquad
T^{\;r}{}_{v}\Big|_{\rm em}=-\frac{Q(v)\dot Q(v)}{4\pi r^{3}}.}
\end{equation}
\rev{A sufficient condition for nonnegative ingoing flux is therefore}
\begin{equation}
\rev{\dot M(v)\ge0,\qquad Q(v)\dot Q(v)\le0,}
\label{eq:WEC-RNAdS-q}
\end{equation}
\rev{corresponding to nondecreasing mass influx and monotone discharge. With these signs, both terms in \eqref{eq:Tvr-split} are nonnegative for $r>0$.}

Let $r_+(v)$ denote the largest positive root of $f\big(r;M(v),Q(v)\big)=0$. Implicit differentiation of $f\big(r_+(v);\cdot\big)=0$ gives
\begin{equation}
\frac{\mathrm{d}r_+}{\mathrm{d}v}
=-\,\frac{\partial_v f}{\partial_r f}\Bigg|_{r_+}
=\frac{\displaystyle \frac{2\dot M}{r_+}-\frac{2Q\dot Q}{r_+^{2}}}
{\displaystyle \frac{2M}{r_+^{2}}-\frac{2Q^{2}}{r_+^{3}}-\frac{2\Lambda}{3}\,r_+ + c\,(3\omega_q+1)\,r_+^{-(3\omega_q+2)}}.
\label{eq:rplus-evolution}
\end{equation}
For a nonextremal outer horizon the denominator is positive. Under the WEC sign choices \eqref{eq:WEC-RNAdS-q} the numerator is nonnegative, hence $\mathrm{d}r_+/\mathrm{d}v\ge 0$. Pure discharge at fixed $M$ gives $\mathrm{d}r_+/\mathrm{d}v>0$ because $Q\dot Q<0$, while pure accretion at fixed $Q$ likewise yields $\mathrm{d}r_+/\mathrm{d}v>0$ through $\dot M>0$. Thus both channels enlarge the outer horizon.

The radius that bounds the negative energy set for a test fragment with opposite sign charge is defined by
\begin{equation}
|q_{1}|\,\Psi\big(v,r_{\mathrm E}(v)\big)=\sqrt{\,f\big(r_{\mathrm E}(v);M(v),Q(v)\big)\,},
\label{eq:rE-def-time}
\end{equation}
with $\Psi=Q(v)/r$. Differentiating \eqref{eq:rE-def-time} gives a closed evolution law
\begin{equation}
\begin{aligned}
\frac{\mathrm{d}r_{\mathrm E}}{\mathrm{d}v}
&=\frac{\displaystyle \frac{|q_{1}|\,\dot Q}{r_{\mathrm E}}
-\frac{1}{2\sqrt{f}}\left(-\frac{2\dot M}{r_{\mathrm E}}+\frac{2Q\dot Q}{r_{\mathrm E}^{2}}\right)}
{\displaystyle \frac{1}{2\sqrt{f}}\left(\frac{2M}{r_{\mathrm E}^{2}}-\frac{2Q^{2}}{r_{\mathrm E}^{3}}-\frac{2\Lambda}{3}\,r_{\mathrm E}
+c(3\omega_q+1)r_{\mathrm E}^{-(3\omega_q+2)}\right)-|q_{1}|\,\Psi_r},\\
\Psi_r&=-\frac{Q}{r_{\mathrm E}^{2}} .
\end{aligned}
\label{eq:rE-evolution-time}
\end{equation}
\rev{The denominator is strictly positive in the exterior under the same monotonicity assumptions used in Section~3: $f_r>0$ and $-|q_{1}|\Psi_r>0$. The numerator contains two competing effects. The explicit electrostatic term $|q_{1}|\dot Q/r_{\mathrm E}\le0$ tends to reduce the electrostatic support of the negative energy boundary, whereas the redshift term proportional to $-\partial_v f/(2\sqrt f)$ can move the boundary outward because monotone discharge lowers the $Q^2/r^2$ part of the lapse. Therefore the sign of $\mathrm{d}r_{\mathrm E}/\mathrm{d}v$ is not universal; it is fixed by the balance in Eq.~\eqref{eq:rE-evolution-time}. For the Maxwell case with $|q_1|_{\rm evol}=1$ used in Figs.~5--6, the direct $Q$ dependence cancels in the defining equation for $r_{\mathrm E}$ at fixed $M$, so pure discharge leaves $r_{\mathrm E}$ nearly stationary while the horizon moves outward. In mixed evolution, mass accretion shifts both $r_+$ and $r_{\mathrm E}$ outward.}

The local bound derived earlier,
\begin{equation}
\eta_{\max}(r_\ast;v)=\frac{|q_{1}|\,\Psi(v,r_\ast)-\sqrt{f\big(r_\ast;M(v),Q(v)\big)}}{E_{0}},
\label{eq:eta-time}
\end{equation}
varies in $v$ through both terms in the numerator. A direct derivative at fixed $r_\ast$ gives
\begin{equation}
\frac{\partial \eta_{\max}}{\partial v}(r_\ast;v)=\frac{1}{E_{0}}
\left[\frac{|q_{1}|\dot Q}{r_\ast}
-\frac{1}{2\sqrt{f}}\left(-\frac{2\dot M}{r_\ast}+\frac{2Q\dot Q}{r_\ast^{2}}\right)\right]_{r=r_\ast}.
\label{eq:deta-dv}
\end{equation}
Under discharge with $\dot Q\le 0$ the first term is nonpositive while the second bracket is nonnegative when $\dot M\ge 0$ and $Q\dot Q\le 0$, so $\partial_v\eta_{\max}$ is typically negative at large $r_\ast$ where the redshift term dominates and can be positive near the horizon where $\sqrt{f}$ is small. \rev{Thus near horizon decays remain the most efficient channel, while the time derivative of the local gain is controlled by a competition between the decreasing electrostatic potential and the decreasing lapse contribution. This is why the evolution of the radial extraction window is parameter and charge dependent rather than universally expansive.}

For explicit illustrations on a finite advanced time interval $0\le v\le v_{\mathrm f}$ we adopt linear evolutions
\begin{equation}
\begin{aligned}
Q(v)&=Q_{\mathrm{ini}}-\gamma v,\qquad \gamma\ge0,\\
M(v)&=
\begin{cases}
M_{0}, & \text{Scenario A: fixed mass, pure discharge},\\[2pt]
M_{0}+\mu v, & \text{Scenario B: accretion with discharge},
\end{cases}
\qquad \mu\ge0 .
\end{aligned}
\label{eq:linear-laws-RN}
\end{equation}
followed by $Q(v)\equiv Q_{\mathrm{ini}}-\gamma v_{\mathrm f}$ and $M(v)\equiv M_{0}+\mu v_{\mathrm f}$ for $v\ge v_{\mathrm f}$. \rev{In Scenario A the apparent horizon expands according to \eqref{eq:rplus-evolution} with $\dot M=0$ and $Q\dot Q<0$. For the representative choice $|q_1|_{\rm evol}=1$, the electrostatic boundary is almost fixed by the $Q$ independent part of the Maxwell defining equation, so the negative energy layer is reshaped mainly by the outward motion of $r_+$. In Scenario B both $r_+$ and $r_{\mathrm E}$ evolve, with $r_+$ increasing due to $\dot M>0$ and $r_{\mathrm E}$ governed by the competition in \eqref{eq:rE-evolution-time}. In both scenarios the conditions \eqref{eq:WEC-RNAdS-q} and \eqref{eq:adiabatic-validity} guarantee a nonnegative ingoing flux within the first order adiabatic approximation. The Kiselev sector is kept fixed during this illustrative evolution. We do not claim that it regularizes the central curvature invariants; the time dependent model is used only to track the outer horizon and the negative energy boundary.}

The operational criterion for successful extraction at fixed $R_{\rm obs}$ remains the pointwise inequality
\begin{equation}
\big(E_{2}-q_{2}\Psi(v,R_{\rm obs})\big)^{2}\ \ge\ f\big(R_{\rm obs};M(v),Q(v)\big),
\label{eq:escape-time}
\end{equation}
to be imposed alongside charge conservation and forward time for both fragments. Combining \eqref{eq:eta-time} with \eqref{eq:escape-time} yields a time dependent sufficient condition identical in form to the static case with the replacements $M\mapsto M(v)$, $Q\mapsto Q(v)$. \rev{As $v$ increases under \eqref{eq:linear-laws-RN}, the feasible domains are updated instantaneously using $M(v)$ and $Q(v)$. Their evolution is controlled by the competing electrostatic and redshift terms in \eqref{eq:deta-dv}; the sign need not be universal over the full parameter space.}

At each $v$ the extremal locus is defined by $f=0=f'$ as in Section~\ref{sec:penrose}. With $Q(v)$ decreasing, the instantaneous extremal charge $Q_{\rm ext}^{2}(r;\Lambda,c,\omega_q)=r^{2}-\Lambda r^{4}+3c\,\omega_q\,r^{\,1-3\omega_q}$ is unchanged, while the working point $(M(v),Q(v))$ moves toward the interior of the black hole domain. Therefore slow discharge drives the geometry away from extremality and enlarges the margin to the threshold, whereas a simultaneous increase of $M(v)$ pushes the curve in the opposite direction. In all cases the instantaneous extraction performance is still governed by the local competition $|q_{1}|\Psi-\sqrt{f}$, so decays as close as permitted to $r_+(v)$ remain optimal throughout the evolution.

Time evolution of the outer apparent horizon $r_{+}(v)$ (solid) and of the generalized electrostatic boundary $r_{\mathrm E}(v)$ (dashed) is shown for all background combinations in Figures 5-6. The shaded strip between the two curves marks the negative energy layer where charged test fragments with opposite sign charge satisfy $E<0$. Figure~5 corresponds to Scenario A with a discharging background $Q(v)=Q_{\mathrm{ini}}-\gamma v$ and fixed mass $M(v)=M_0$, while Figure~6 depicts Scenario B where discharge proceeds together with linear accretion $M(v)=M_0+\mu v$.

\rev{In Scenario A the horizon grows moderately with $v$ because $Q\dot Q<0$ reduces the repulsive term in $f$ and shifts the largest root outward. For $|q_1|_{\rm evol}=1$, the Maxwell ergosphere equation cancels the explicit $Q^2/r^2$ contribution and the electrostatic boundary remains nearly stationary at fixed $M$. Consequently the shaded layer generally narrows slowly as the horizon approaches the boundary. The largest layers still occur for the larger mass set $M=1.2$ and for the more negative state parameter $\omega_q=-\tfrac{2}{3}$, which both lower the lapse in the exterior and push the $E<0$ domain to larger radii. A less negative cosmological constant $\Lambda=-0.02$ changes the fixed finite radius scale and leads to a milder variation of the horizon and boundary separation over the displayed interval. Across panels, increasing $c$ shifts the electrostatic boundary upward, reflecting the quintessence reduction of the lapse and the associated enlargement of the kinematically allowed region.}

\rev{In Scenario B the mass growth $\dot M=\mu>0$ amplifies the outward motion of the horizon and also shifts $r_{\mathrm E}$ outward. The net effect depends on the balance between $\mu$ and $\gamma$: for the parameter choices used in the plots the layer persists, but its thickness changes more slowly than the individual radii because both curves move in the same direction.} \rev{The parametric ordering remains consistent with the static trends established earlier. Larger $M$ or $Q$ and larger $c$ support a thicker negative energy shell at a fixed time, while a more negative $\Lambda$ affects the finite radius behavior through the stronger AdS barrier. The difference between $\omega_q=-\tfrac{2}{3}$ and $\omega_q=-\tfrac{1}{2}$ is uniform across backgrounds: the former yields a larger electrostatic boundary and a wider allowed layer at fixed $v$.}

\rev{Figs.~5--6 together provide a time resolved map of how discharge and accretion reshape the window for charged Penrose extraction. In both scenarios the most favorable configuration remains a near horizon decay, but the accessible interval in $r$ is history dependent: pure discharge mainly moves the horizon at fixed $r_{\mathrm E}$ for $|q_1|_{\rm evol}=1$, whereas simultaneous accretion moves both radii and can partially preserve the layer thickness without reversing the ordering across $(M,Q,\Lambda,c,\omega_q)$.}

\begin{figure}[H]
  \centering
  \includegraphics[width=\textwidth]{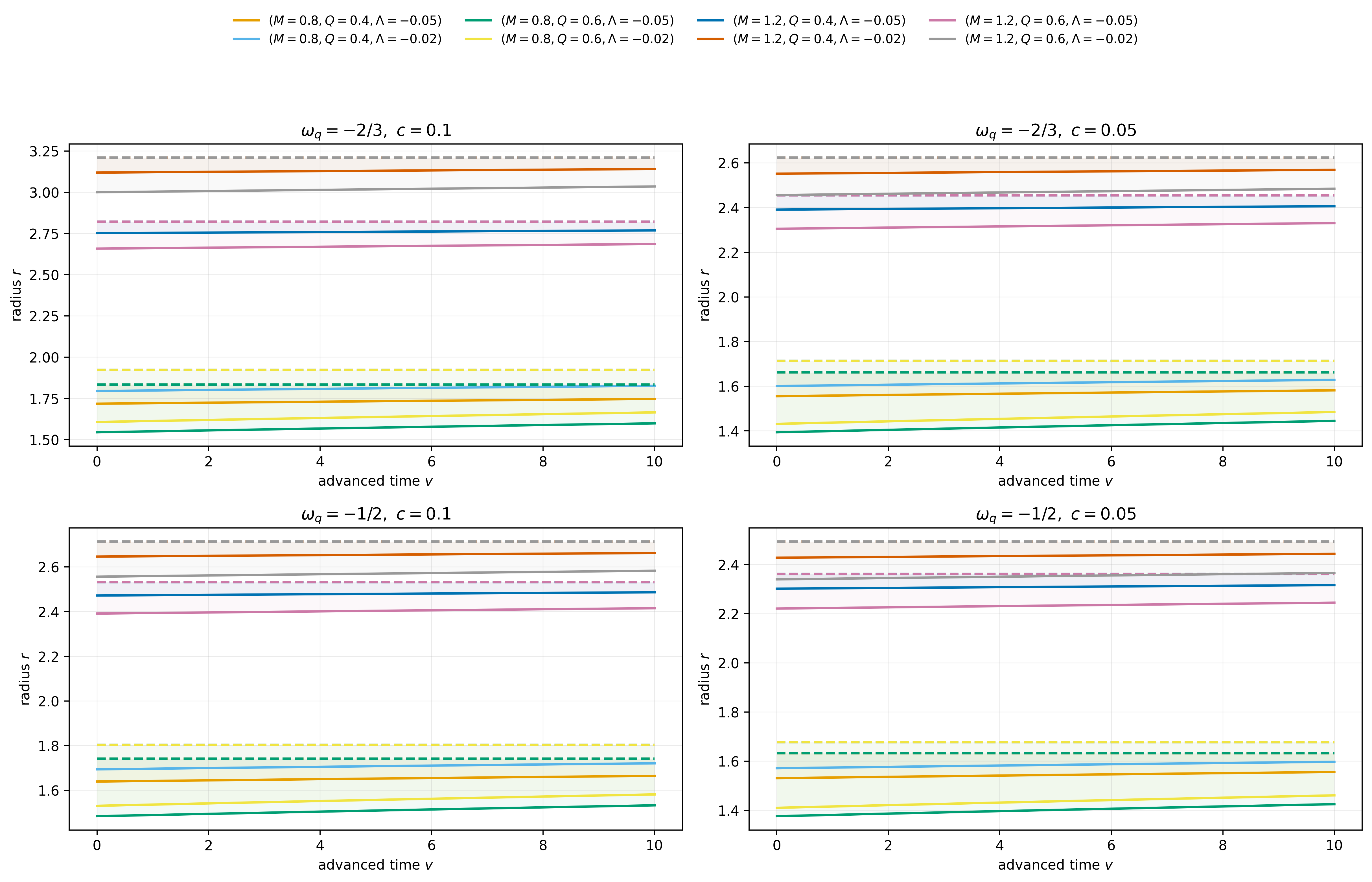}
  \caption{\rev{Evolution of the outer horizon $r_{+}(v)$ (solid) and the generalized electrostatic boundary $r_{\mathrm E}(v)$ (dashed) for Scenario A with $M(v)=M_0$, $Q(v)=Q_{\mathrm{ini}}-\gamma v$, $\gamma=5\times10^{-3}$, and $|q_1|_{\rm evol}=1$ over $0\le v\le10$. Each panel fixes $(\omega_q,c)$ and shows all $(M,Q,\Lambda)$ combinations in Table~\ref{tab:numerics}. The shaded strip indicates the negative energy layer $r_{+}<r<r_{\mathrm E}$.}}
  \label{fig:evol_rplus_rE_A}
\end{figure}

\begin{figure}[H]
  \centering
  \includegraphics[width=\textwidth]{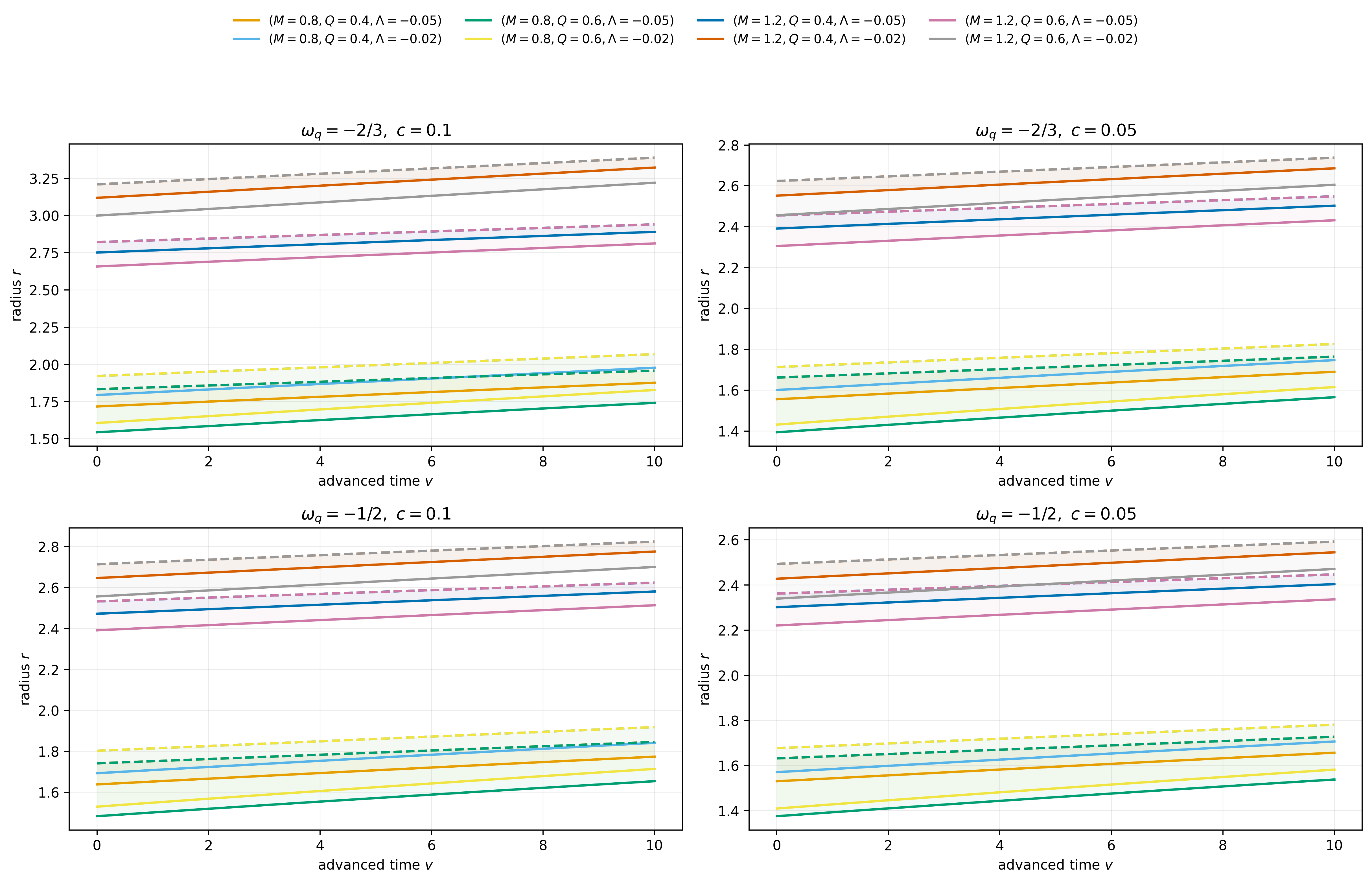}
  \caption{\rev{Same as Fig.~\ref{fig:evol_rplus_rE_A} for Scenario B with $M(v)=M_0+\mu v$, $Q(v)=Q_{\mathrm{ini}}-\gamma v$, $\mu=\gamma=5\times10^{-3}$, and $|q_1|_{\rm evol}=1$ over $0\le v\le10$. Accretion moves both $r_{+}$ and $r_{\mathrm E}$ outward, so the layer thickness is governed by their relative drift.}}
  \label{fig:evol_rplus_rE_B}
\end{figure}
\FloatBarrier

\section{Conclusions and outlook}

\rev{ In this paper, we develop the analysis into a controlled framework for charged particle kinematics and Penrose extraction in RN--AdS geometries surrounded by a Kiselev type anisotropic dark energy sector. The generalized electrostatic ergosphere is determined by $|q|\Psi(r_E) = \sqrt{f(r_E)}$. Under the explicitly stated assumptions $f'(r) > 0$ and $\Psi'(r) < 0$ in the exterior domain, this boundary is unique and the turning point of a negative energy trajectory is also unique. The result implies capture of the negative energy fragment within the test particle approximation.}

\rev{For the charged Penrose process we derive a local upper bound on the escaping energy and combine it with a finite radius reception criterion appropriate to AdS asymptotics. The figures separate the continuous charge dependence from the multi-parameter survey: Figs.~2 and 3 display shaded allowed domains and continuous efficiency maps, while Fig.~4 presents the full feasibility map for reception at $R_{\rm obs}=10$. }

\rev{The time dependent part has also been clarified. The evolution laws for $r_+(v)$ and $r_E(v)$ are first order adiabatic tracking relations for slowly varying $M(v)$ and $Q(v)$. The validity conditions are $E_i/M\ll1$, $|q_i|/|Q|\ll1$, and $\epsilon_{\rm ad}\ll1$. Within this regime, monotone discharge and accretion reshape the negative energy layer through different terms in Eq.~\eqref{eq:rE-evolution-time}; for the representative Maxwell choice $|q_1|_{\rm evol}=1$, pure discharge mainly shifts the horizon while mixed accretion moves both radii. }

\rev{The main trends are robust within the plotted parameter domain: larger $M$, larger $Q$, larger $c$, and the more negative value of $\omega_q$ increase the size of the negative energy layer and raise the local efficiency bound, while the AdS scale controls finite radius reception. Near extremality enhances the local kinematic ceiling, but the discussion emphasizes that horizon instabilities, quantum discharge, radiative losses, and fragment stress energy can regulate this idealized behavior. Future work should therefore incorporate rotation, dynamical dark energy, nonlinear electrodynamics beyond monotone profiles, and self-consistent backreaction.}

\end{document}